\documentclass[
    sort&compress
    ,final            
    ,numberedheadings 
  ]
  {aipproc}

\layoutstyle{8x11single}

\begin{document}

  \newcommand {\nc} {\newcommand}
  \nc {\beq} {\begin{eqnarray}}
  \nc {\eeq} {\nonumber \end{eqnarray}}
  \nc {\eeqn}[1] {\label {#1} \end{eqnarray}}
  \nc {\eol} {\nonumber \\}
  \nc {\eoln}[1] {\label {#1} \\}
  \nc {\ve} [1] {\mbox{\boldmath $#1$}}
  \nc {\mrm} [1] {\mathrm{#1}}
  \nc {\half} {\mbox{$\frac{1}{2}$}}
  \nc {\thal} {\mbox{$\frac{3}{2}$}}
  \nc {\fial} {\mbox{$\frac{5}{2}$}}
  \nc {\la} {\mbox{$\langle$}}
  \nc {\ra} {\mbox{$\rangle$}}
  \nc {\etal} {\emph{et al.\ }}
  \nc {\eq} [1] {(\ref{#1})}
  \nc {\Eq} [1] {Eq.~(\ref{#1})}
  \nc {\Ref} [1] {Ref.~\cite{#1}}
  \nc {\Refc} [2] {Refs.~\cite[#1]{#2}}
  \nc {\Sec} [1] {Sec.~\ref{#1}}
  \nc {\chap} [1] {Chapter~\ref{#1}}
  \nc {\anx} [1] {Appendix~\ref{#1}}
  \nc {\tbl} [1] {Table~\ref{#1}}
  \nc {\fig} [1] {Fig.~\ref{#1}}
  \nc {\bfig} {\begin{figure}}
  \nc {\efig} {\end{figure}}
  \nc {\ex} [1] {$^{#1}$}
  \nc {\Sch} {Schr\"odinger }
  \nc {\flim} [2] {\mathop{\longrightarrow}\limits_{{#1}\rightarrow{#2}}}

\title{Time-dependent analysis of the nuclear and Coulomb dissociation
of $^{11}$Be}


\author{Pierre Capel}{
  address={TRIUMF, 4004 Wesbrook Mall, Vancouver, B.C., Canada V6T 2A3}
}

\author{G\'erald Goldstein}{
  address={Physique Quantique, C.P. 165/82 and
Physique Nucl\'eaire Th\'eorique et Physique Math\'ematique, C.P. 229,
Universit\'e Libre de Bruxelles, B-1050 Brussels, Belgium}
}

\author{Daniel Baye}{
  address={Physique Quantique, C.P. 165/82 and
Physique Nucl\'eaire Th\'eorique et Physique Math\'ematique, C.P. 229,
Universit\'e Libre de Bruxelles, B-1050 Brussels, Belgium}
}

\begin{abstract}
The breakup of $^{11}$Be on carbon and lead targets around
70 MeV/nucleon is investigated within a semiclassical framework.
The role of the $\frac{5}{2}^+$ resonance is analyzed in both cases.
It induces a narrow peak in the
nuclear-induced breakup cross section, while its
effect on Coulomb breakup is small.
The nuclear interactions between the projectile and the target
is responsible for the transition toward this resonant state.
The influence of the parametrization of the \ex{10}Be-n
potential that simulates \ex{11}Be is also addressed.
The breakup calculation is found to be dependent
on the potential choice.
This leads us to question the reliability of this technique
to extract spectroscopic factors.
\end{abstract}

\maketitle

\section{Introduction}

The \ex{11}Be nucleus is one of the best known one-neutron halo nuclei.
Its halo structure has thus been the subject of many
theoretical and experimental analyzes \cite{Tan96}.
In particular, breakup reactions are used as tools to
extract its structure properties \cite{Nak94,Fuk04}.
Various theoretical models have been developed to
interpret the experimental data \cite{AN03}: perturbation expansion,
adiabatic approximation \cite{Jmsu05a},
Eikonal model \cite{Tosmsu05a},
coupled channel with a discretized continuum
(CDCC) \cite{Tmsu05a,Kmsu05a},
and numerical resolution of a three-dimensional
time-dependent \Sch equation \cite{Bmsu05a}.

Recently, the breakup of \ex{11}Be on both lead and carbon targets
has been measured at RIKEN around 70~MeV/nucleon \cite{Fuk04}.
In the present talk, we investigate these reactions
with a time-dependent technique.
This reaction model is based on a semiclassical approximation
\cite{AW75,Bmsu05a} in which the relative motion of the projectile
and the target is approximated by a classical trajectory.
Therefore, the projectile is seen as evolving in a time-dependent
potential that simulates its interaction with the target.
This approximation leads to the resolution of a time-dependent
\Sch equation. Different techniques have been developed to solve
this equation \cite{KYS94,EBB95,MB99,TS01r,Fal02,CBM03c}.
We use the technique described in \Ref{CBM03c}.

Up to now, \ex{11}Be is described in all reaction models
as a two-body system: a halo neutron loosely bound to a structureless
\ex{10}Be core. The interaction between the neutron and the
core is modeled by a simple local potential.
This \ex{10}Be-n potential is usually adjusted to reproduce the
bound states of \ex{11}Be \cite{KYS94,EBB95,MB99,TS01r,Fal02}.
It is of course important to analyze the accuracy of that description.
In particular, one needs to know what is reproducible using
such a simple model, and what is not.
In a recent paper, we studied the breakup of \ex{11}Be
on a \ex{12}C target \cite{CGB04}.
For that study, we developed a new \ex{10}Be-n potential
that reproduces not only the bound states of \ex{11}Be, but
also its first resonant state above the one-neutron threshold.
That resonance is found to induce a narrow peak in the breakup
cross section. A similar peak is observed in the experimental data.
This suggests that the resonance can be fairly well reproduced
in the two-body description,
and that its presence in reaction models is required
to reproduce the experimental data.
In this talk, we present the results of this analysis.
We also present recent calculations
of the Coulomb breakup of \ex{11}Be. In particular,
we discuss the role played by the resonance in that reaction
and compare it to its role played in the dissociation on \ex{12}C.

Besides the capability of this simple two-body description
to reproduce physical levels of \ex{11}Be, the sensitivity
of the calculations to the parametrization of the \ex{10}Be-n
potential must also be assessed.
In particular, the breakup cross section should not
be too sensitive to the potential choice
if one wishes to reliably extract
spectroscopic information from measurements.
In this talk, we present the first results of such an analysis.
The results of calculations
of the Coulomb breakup of \ex{11}Be performed
with different \ex{10}Be-n potentials are discussed.

The talk is structured as follows.
After a brief description of the time-dependent model
and the parametrizations of the potential that describes \ex{11}Be,
we present, in \Sec{BeC},
the results we have obtained in the breakup on \ex{12}C \cite{CGB04}.
The analysis of the Coulomb breakup of \ex{11}Be is discussed
in \Sec{BePb}.
The final section contains our concluding remarks.

\section{Theoretical framework}

\subsection{Time-dependent model}

\begin{figure}
  \includegraphics[height=7cm]{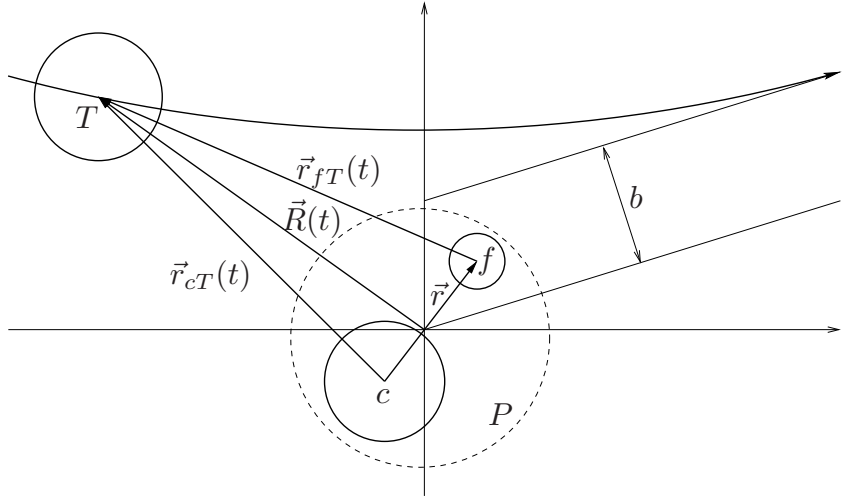}
  \caption{Semiclassical scheme of the reaction.
In the projectile ($P$) rest frame, the target ($T$)
follows a classical trajectory described by the
time-dependent relative coordinate $\vec{R}$.
The projectile is assumed to have
a two-body structure: a fragment $f$ loosely bound
to a core $c$. Their relative coordinate is $\vec{r}$.
}\label{f0}
\end{figure}

We consider the breakup of a projectile $P$ by a target $T$.
The projectile $P$ is assumed to have a two-body structure:
a pointlike and structureless fragment $f$
(of mass $m_f$ and charge $Z_fe$) loosely
bound to a structureless core $c$ (of mass $m_c$ and charge $Z_ce$).
The target is seen as a structureless particle of mass $m_T$
and charge $Z_Te$.
In the semiclassical approximation \cite{AW75,Bmsu05a},
the $P$-$T$ relative motion is treated classically:
in the projectile rest frame,
the target is assumed to follow a classical trajectory (see \fig{f0}).
Therefore, the interaction between the projectile and the target is
simulated by a time-dependent potential.
The internal motion of the projectile, however,
is treated quantum mechanically. The wave function $\Psi$
describing this motion is solution of the following
time-dependent \Sch equation:
\beq
i\hbar\frac{\partial}{\partial t}\Psi(\vec{r},t)
&=&[H_0(\vec{r})+V(\vec{r},t)]\Psi(\vec{r},t)\nonumber\\
&=&\left\{-\frac{\hbar^2}{2\mu}\Delta+V_{cf}(r)
+V_{cT}[r_{cT}(\vec{r},t)]+V_{fT}[r_{fT}(\vec{r},t)]
-\frac{(Z_c+Z_f)Z_T e^2}{R(t)}\right\}\Psi(\vec{r},t),
\eeqn{e0}
where $\vec{r}$ is the relative coordinate of the fragment
to the core, $\vec{R}$ is the time-dependent coordinate
that describes the trajectory of the target in the projectile
rest frame, and $\vec{r}_{cT}$ and $\vec{r}_{fT}$ are respectively
the core-target and fragment-target relative coordinates as
illustrated in \fig{f0}.

In \Eq{e0}, the Hamiltonian $H_0$ describes the internal structure of the
two-body projectile. It is the sum of the kinetic term and the
local potential $V_{cf}$, which simulates the interaction between
the core and the fragment (see \Sec{Be11}).
The potentials $V_{cT}$ and $V_{fT}$ model the core-target
and fragment-target interactions, respectively.
They comprise a Coulomb term and a short-range optical potential,
which simulates the nuclear interaction.
The latter is usually chosen in the literature.

\Eq{e0} is solved with the initial condition that
at time $t\rightarrow -\infty$ the projectile is in
its ground state.
The wave function $\Psi$ at time $t$ is then obtained
iteratively using the evolution algorithm
described in \Ref{CBM03c}.
The calculation is performed for different trajectories
parametrized by the impact parameter $b$.
For each trajectory, we deduce the breakup probability
by projecting the output wave function
$\Psi(\vec{r},t\rightarrow +\infty)$ onto the positive eigenstates
of $H_0$ that describe the continuum of the projectile.
The breakup cross section is then obtained by summing this probability
over all impact parameters.

\subsection{\ex{11}Be description}\label{Be11}
As done in previous works \cite{KYS94,EBB95,MB99,TS01r,Fal02,CBM03c,CGB04},
we describe \ex{11}Be as a neutron loosely bound to a \ex{10}Be core.
The \ex{10}Be core is assumed to be in its $0^+$ ground state, and
the spectroscopic factor associated to that configuration
is set equal to unity.
The potential which simulates the \ex{10}Be-n interaction
is composed of a central part plus a spin-orbit coupling term
\beq
V_{cf}(r)=V_0(r)+\vec{L}\cdot\vec{I}V_{LI}(r),
\eeqn{e3}
where $\vec{L}$ is the orbital momentum of the \ex{10}Be-n
relative motion, and $\vec{I}$ is the spin of the neutron.
The central part of $V_{cf}$ has a Woods-Saxon form factor
\beq
V_0(r)=-V_l f(r,R_0,a),
\eeqn{e10}
where
\beq
f(r,R_0,a)=\left[1+\exp\left(\frac{r-R_0}{a}\right)\right]^{-1}.
\eeqn{e11}
The spin-orbit coupling term
has the usual Thomas form factor
\beq
V_{LI}(\ve{r})=V_{LS}\frac{1}{r}\frac{d}{dr}f(r,R_0,a).
\eeqn{e12}
The radius of the form factor is parametrized as usual:
$R_0=r_0A_c^{1/3}$.

The depths of the potential are adjusted to reproduce
the energies of the low-lying states of \ex{11}Be.
The well known shell inversion observed between the
bound states is reproduced by using a parity-dependent
depth of the central part of the potential $V_l$.
The $\half^+$ ground state is modeled by a $1s1/2$ state,
the $\half^-$ excited state by a $0p1/2$ state,
and the first $\fial^+$ resonance is reproduced in the $d5/2$ wave.

\begin{table}
\begin{tabular}{c c c c c c} \hline
  \tablehead{1}{c}{b}{Potential} &
  \tablehead{1}{c}{b}{$V_{l \mrm{even}}$} &
  \tablehead{1}{c}{b}{$V_{l \mrm{odd}}$} &
  \tablehead{1}{c}{b}{$V_{LS}$} &
  \tablehead{1}{c}{b}{$a$} &
  \tablehead{1}{c}{b}{$r_0$}\\
 & (MeV) & (MeV) & (MeV fm\ex{2}) & (fm) & (fm)\\ \hline
V1 & 62.52 & 39.74 & 21.0 & 0.6 & 1.2\\
V2 & 66.325 & 38.37 & 12.44 & 0.5 & 1.2\\
V3 & 58.905 & 40.025 & 27.68 & 0.7 & 1.2\\
V4 & 71.28 & 49.015 & 29.95 & 0.6 & 1.1\\
V5 & 55.25 & 32.515 & 12.86 & 0.6 & 1.3\\
V6 & 59.05 & 59.05 & 0 & 0.62 & 1.236\\ \hline
\end{tabular}
\caption{Parameters of the $^{10}$Be-n potentials
[see Eqs. \protect\eq{e10}-\protect\eq{e12}].
Note that $R_0$ used in \protect\eq{e10}-\protect\eq{e12}
is parametrized as $r_0 A_c^{1/3}$.}
\label{t1}
\end{table}

In order to study the sensitivity of our calculations to the
potential choice, we developed five sets of parameters
that reproduce the physical states mentioned above.
They are summarized in \tbl{t1}.
The first potential (V1) has been devised for our recent calculation
of the breakup of \ex{11}Be on \ex{12}C \cite{CGB04}
(see also \Sec{BeC}).
The next four (V2 to V5) have been obtained by varying either
the diffuseness or the radius of the Woods-Saxon form factor.
The values were chosen to encompass those used by most
other groups \cite{KYS94,EBB95,TS01r,Fal02}.
Besides the three physical levels, these potentials all
exhibit two unphysical bound states: $0s1/2$ and $0p3/2$.
These states correspond to the shells occupied by the neutrons in the core
and are forbidden by the Pauli principle.
Their energies have not been adjusted, and thus vary from one
potential to the other.
Each potential also displays a $d3/2$ resonance.
This resonance does not correspond to any known physical state.
Therefore it has not been fitted. Its location and width vary
with the potential choice.
Since it is very broad and located at high energy,
we doubt this resonance might play any significant role in our calculations.
In \tbl{t1}, we also list a sixth potential (V6)
developed by Fukuda \etal \cite{Fuk04}.
It reproduces only the ground state energy and
does not contain a spin-orbit coupling term.

\section{Breakup of \ex{11}Be on \ex{12}C}\label{BeC}
\subsection{Breakup cross section}\label{BeC.bu}

\begin{figure}
  \includegraphics[height=6cm]{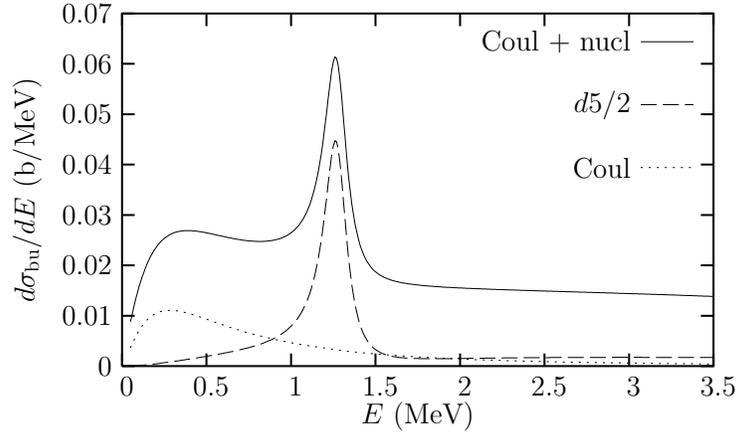}
  \caption{Breakup cross section
of \ex{11}Be on \ex{12}C at 67~MeV/nucleon
as a function of the \ex{10}Be-n relative energy $E$ after breakup.
The full line corresponds to the
calculation considering both
the nuclear and Coulomb interactions between the
projectile and the target.
The dashed line displays its $d5/2$ component, responsible for the peak
at the resonance energy.
The result obtained with a pure Coulomb $P$-$T$ interaction
is displayed as the dotted line.
}\label{f1}
\end{figure}

Recently, the breakup of \ex{11}Be on \ex{12}C at 67~MeV/nucleon
has been measured at RIKEN \cite{Fuk04}.
We analyze this reaction within the semiclassical framework
described in the previous section.
In \fig{f1}, the breakup cross section is displayed as a
function of the relative energy $E$ between the \ex{10}Be core
and the neutron after breakup.
The full line corresponds to the results obtained considering both
nuclear and Coulomb interactions between the projectile and the target.
The conditions of the calculations are those described in \Ref{CGB04}.
In particular, \ex{11}Be is described by the \ex{10}Be-n
potential V1 of \tbl{t1} developed in that previous work.
For the \ex{10}Be-\ex{12}C interaction, we first use
the optical potential developed by Al-Khalili, Tostevin and Brook,
which has been adjusted to reproduce
\ex{10}Be-\ex{12}C scattering data \cite{ATB97}.
The n-\ex{12}C interaction is simulated by the
Becchetti and Greenlees parametrization \cite{BG69}.
For comparison, we also display the cross section computed
with a pure Coulomb interaction between the projectile and
the target (dotted line).
This emphasizes the strong dominance of the nuclear
interactions in this dissociation reaction.

First, the breakup cross section is significantly enhanced
when optical potentials are considered.
This is true on the entire energy range, but is particularly
striking at high energy.
The presence of nuclear $P$-$T$ interactions leads to
a gentle decrease of the cross section with energy,
while a pure Coulomb interaction induces a rapid drop
of the cross section beyond 0.5~MeV.

Second, a narrow peak is observed in the breakup cross
section obtained with optical potentials.
This peak is due to the $d5/2$ resonance
present in our description of \ex{11}Be.
It is indeed located at the same energy and exhibits the same width as
that resonance. Moreover, it appears solely in the contribution
of the $d5/2$ partial wave
to the cross section (dashed line).
The absence of peak in the purely Coulomb result
indicates that $P$-$T$ nuclear interactions are necessary
to populate that resonant state.

\subsection{Comparison with experiment and analysis of the
influence of the optical potentials}

\begin{figure}
  \includegraphics[height=6cm]{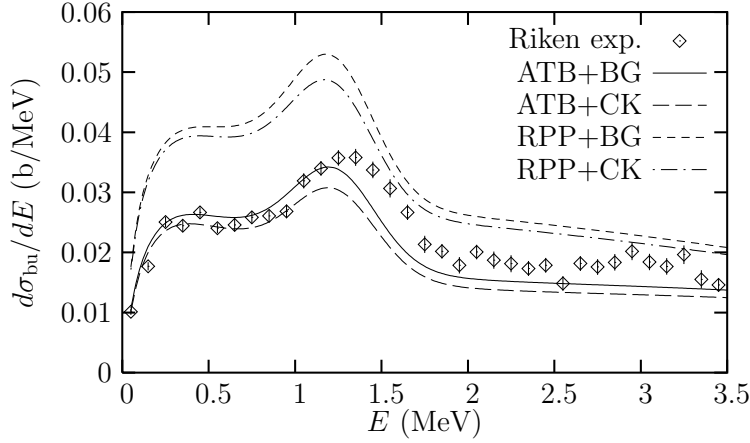}
  \caption{Theoretical and experimental
breakup cross sections
of \ex{11}Be on \ex{12}C as a function of energy.
The four curves correspond to the calculations
performed with various combinations of optical potentials
simulating the nuclear $P$-$T$ interactions.
The theoretical results have been convoluted with energy
resolution.
Experimental data are from \Ref{Fuk04}.
}\label{f2}
\end{figure}

In \fig{f2}, we compare the results of our calculation
with the breakup cross section measured at RIKEN \cite{Fuk04}.
The full line (labeled ATB+BG) corresponds to the
full line of \fig{f1} convoluted with the experimental
energy resolution.
The main effect of this convolution is to significantly broaden
the resonance peak and slightly shift it toward lower energies.

We observe a very good agreement between theory and experiment.
Note that all the parameters have been fixed prior to the calculation;
there is no adjustment of our results to the experimental data.
At low energy, theory and experiment exhibit the same behavior.
In particular, they both display a peak in the vicinity of the
$\fial^+$ resonance. Moreover, these peaks have approximately the same
shape (height and width). This confirms that the low-lying
resonance in the \ex{11}Be spectrum has a significant influence
on the nuclear induced breakup.

The discrepancy between theory and experiment observed
at larger energies is most likely due to the fact
that our \ex{11}Be model does not reproduce any known physical state
above the $\fial^+$ one. The experimental \ex{11}Be spectrum indeed
includes two other low-lying resonances, which should
have an influence on the breakup as well.
The first is located at 2.2~MeV and is probably responsible
for the underestimation of the theory with respect
to the experiment.
The second is located at 2.9~MeV and may explain the small
peak observed in the RIKEN data.

With the aim of analyzing the influence of the parametrization
of the nuclear $P$-$T$ interactions on those results,
we perform the same calculation using different
sets of optical potentials.
Besides the potential of Al-Khalili, Tostevin and Brook
\cite{ATB97} (labeled ATB), we choose another potential
to simulate the \ex{10}Be-\ex{12}C nuclear interaction.
Following Chatterjee \cite{Chat03}, we use a parametrization
listed in the Perey and Perey compilation \cite{PP76}
which is a simplified expression of a potential
developed by Robson \cite{Rob71} to reproduce the
scattering of \ex{10}B on \ex{12}C at 18~MeV (labeled RPP).
As an alternative to the Becchetti and Greenlees potential (BG) \cite{BG69}
for simulating the n-\ex{12}C interaction, we consider
the potential developed by Comfort and Karp (CK) to
reproduce scattering data of protons impinging on \ex{12}C \cite{CK80}.

The breakup cross sections obtained with the
four possible combinations of those potentials
are displayed in \fig{f2} after convolution with the energy resolution.
All curves exhibit the same pattern. In particular, they
all display similar peaks near the $\fial^+$ resonance energy.
This result shows that the optical potential choice has but little
influence on the shape of that peak.
It therefore confirms that the peak reflects the presence
of the low-lying resonance in our \ex{11}Be model.

The main difference between the four
calculations is due to the \ex{10}Be-\ex{12}C potential.
The amplitude of the breakup cross section is indeed
multiplied by almost 2 when the ATB potential is substituted by
the RPP parametrization.
This increase is due to the much smaller imaginary part of RPP.
On the other hand, it seems that both n-\ex{12}C interactions are
equivalent to describing breakup reactions.
The difference between the cross sections obtained with
the BG and CK potentials is indeed rather small.

A detailed analysis of the breakup probability
as a function of the impact parameter $b$
confirms these results (see \Ref{CGB04}).
In particular, it shows that the internal
structure of the projectile, like the presence of the
$\fial^+$ resonance, is probed only when the
nuclear $P$-$T$ interactions are taken into account.

These results show that, as expected, the nuclear $P$-$T$ interactions
play a dominant role in the dissociation of halo nuclei on light targets.
In particular, these interactions emphasize the presence
of low-lying resonances in the projectile spectrum.
These resonances must therefore be taken into account
in order to reproduce the experimental data.
This suggests that nuclear induced breakup can be used as
a probe of the continuum spectrum of the projectile.

\section{Breakup of \ex{11}Be on \ex{208}Pb}\label{BePb}
\subsection{Breakup cross section}\label{BePb.bu}

\begin{figure}
  \includegraphics[height=7cm]{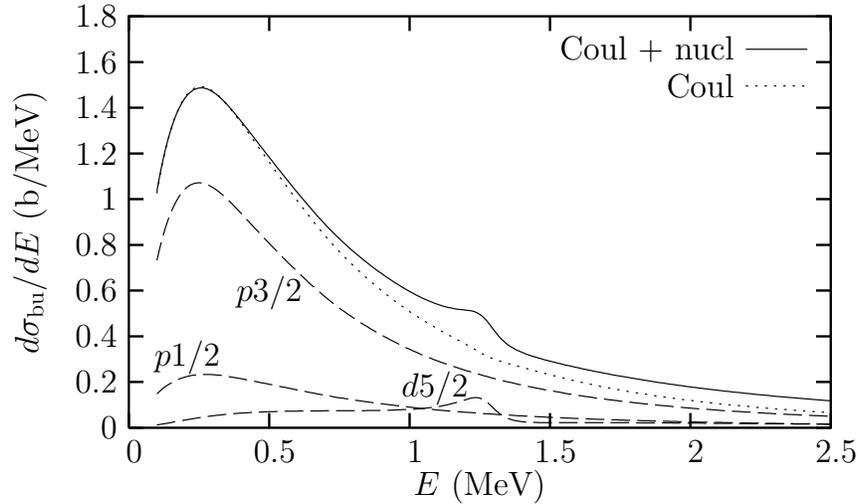}
  \caption{Breakup cross section of \ex{11}Be on \ex{208}Pb
at 69~MeV/nucleon as a function of energy.
The full line corresponds to the calculation
performed with both Coulomb and nuclear $P$-$T$ interactions
Some major contributions of the partial waves to the cross section
are displayed as dashed lines.
The dotted line
corresponds to the result obtained with a pure Coulomb $P$-$T$ interaction.
}\label{f3}
\end{figure}

We now turn to the Coulomb breakup of \ex{11}Be on \ex{208}Pb.
This reaction has been recently remeasured at RIKEN
at 69~MeV/nucleon \cite{Fuk04}. Moreover, it is interesting to
see how the $\fial^+$ resonance in the \ex{11}Be spectrum
will affect the dissociation reaction when it is Coulomb dominated.
The breakup cross section is displayed in \fig{f3} as a
function of the energy $E$.
The calculations were performed with the same conditions as in
\Ref{CBM03c}, but for the \ex{10}Be-n potential, which is the
same as in \Sec{BeC.bu} (i.e. V1 of \tbl{t1}).

The full line corresponds to the calculation performed including both
Coulomb and nuclear $P$-$T$ interactions.
In this case, the optical potential simulating
the nuclear interaction between \ex{10}Be and \ex{208}Pb
is adapted from an $\alpha$-\ex{208}Pb potential \cite{Bon85}
as explained in \Ref{CBM03c}.
The n-\ex{208}Pb potential is chosen to be the
Becchetti and Greenlees parametrization \cite{BG69}.
The cross section obtained with a purely Coulomb potential
between \ex{11}Be and \ex{208}Pb is displayed as a dotted line.
In that case, the nuclear interactions are simulated by an
impact parameter parameter cutoff.

As expected, the breakup of \ex{11}Be on \ex{208}Pb is
strongly dominated by the Coulomb interaction.
The discrepancy between the cross sections computed
with and without the nuclear $P$-$T$ interactions is
indeed small.
However, there remain some interesting differences.
As in the breakup on \ex{12}C, but to a much smaller extent,
the use of optical potentials leads to an
increase of the breakup cross section at high energy.
This has already been observed by Typel and Shyam \cite{TS01r}.
As explained in \Ref{CBM03c}, the
effect of the nuclear interactions, though small,
cannot be fully reproduced by a mere impact parameter cutoff.
We also observe that the nuclear $P$-$T$ interactions induce
a small bump in the breakup cross section.
As in the previous case, this bump is due to the presence
of the $\fial^+$ state in our description of \ex{11}Be:
it is located at the resonance energy
and is due only to the contribution of the $d5/2$ partial wave
(lowest dashed line in \fig{f3}).
It is much smaller than in the previous case (cf. \fig{f1}).
Although the Coulomb field is very strong in this case,
it appears only when optical potentials are used.
This effect confirms that only the nuclear interactions can
significantly populate the $\fial^+$ resonant state.
Therefore, this low-lying resonance
does not affect much the Coulomb dissociation
of \ex{11}Be.

Note that, since the breakup of \ex{11}Be on \ex{208}Pb
is Coulomb dominated, the calculation of its cross section
is much less sensitive to the optical potential choice
than in the nuclear induced breakup.
A variation of 20\% in the amplitude of the optical potentials
leads to only 2\% variation in the breakup cross section.

\subsection{Comparison with experiment and analysis of the
sensitivity to the \ex{10}Be-n potentials}

\begin{figure}
  \includegraphics[height=7cm]{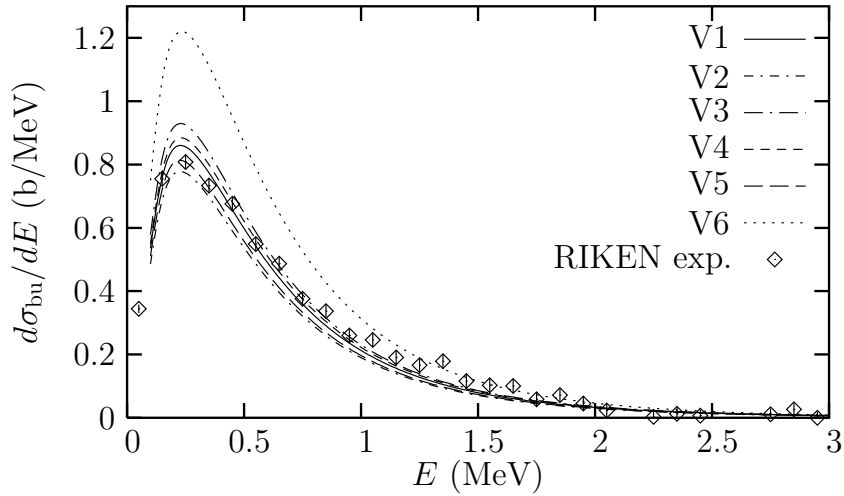}
  \caption{Breakup cross sections of \ex{11}Be on \ex{208}Pb
at 69~MeV/nucleon obtained for $b>30$~fm.
The six curves correspond to calculations performed with
the different potentials of \tbl{t1}.
Experimental data are from \Ref{Fuk04}.
}\label{f4}
\end{figure}

When we compare the results of our calculations with the
experimental breakup cross sections measured at RIKEN
\cite{Fuk04}, we find a rather good agreement
between theory and experiment.
This is illustrated in \fig{f4}, where the Coulomb breakup
cross section corresponding to $b>30$~fm is plotted as a
function of the energy.
At these impact parameters, the nuclear interactions between
projectile and target are
completely negligible. This enables us to get rid of the
problem of their simulation.
The cross section obtained from the calculation
presented in the previous section
(i.e. using potential V1 of \tbl{t1})
is displayed by the full line.
This result is indeed very close to the experiment.
Note that no parameter has been adjusted to fit the data.
In particular, the theoretical cross section has not been
scaled by any factor. This suggests that \ex{11}Be
is well described by a neutron loosely bound to a
\ex{10}Be core in its $0^+$ ground state.
The spectroscopic factor of that configuration
should therefore be close to unity.

With the aim of testing the sensitivity of our results
to the \ex{10}Be-n potential, we perform the same
calculation with different potentials.
These potentials are obtained by varying either
the radius or the diffuseness of the Woods-Saxon
form factor (see \Sec{Be11}).
They are the potentials V2 to V5 given in \tbl{t1}.
The corresponding breakup cross sections are displayed
in \fig{f4}.
All curves exhibit the same shape. The only difference
lies in their amplitude, which varies by about 15\%.
Surprisingly, these variations are not due
to the asymptotic normalization constant of the initial ground state.
For example, V4 leads to a larger breakup cross section
than V5 (by approximately 9\%), although its ANC (0.82) is
smaller than that of V5 (0.87).
This puzzling feature is currently under investigation.
Up to now, it seems that this difference is due to the
scattering properties of the potentials (e.g. scattering length),
which differ from one potential to the other.
Nevertheless, all these results confirm that the
spectroscopic factor of the \ex{10}Be($0^+$)-n configuration
should be close to 1.

For the analysis of their measurements, Fukuda \etal
use another \ex{10}Be-n potential (V6 in \tbl{t1}) \cite{Fuk04}.
It is a Woods-Saxon potential whose depth is
adjusted only to the ground state energy of \ex{11}Be.
It does not include any spin-orbit coupling term.
From that analysis, they deduce a spectroscopic
factor of 0.7, much lower than what we get from our calculations.
In order to understand the discrepancy between our value
and theirs, we perform a calculation using potential V6
within our model. The corresponding cross section
is displayed as a dotted line in \fig{f4}.
It lies approximately 30\% above the V1 curve,
which explains the lower spectroscopic factor.

The reason for this difference is still
to be analyzed. However, the discrepancy
obtained within the same reaction model
using different $V_{cf}$ supposed to describe
the same nucleus is very large.
Therefore, we wonder whether this technique is
reliable for extracting spectroscopic factors.
This result indicates that a strong effort
should be made to improve the description
of a halo nucleus used in reaction models.
At least, the core-fragment potential should
be constrained by other experimental data
or predictions from precise structure models.

\subsection{Influence of Pauli-forbidden states}

In the preceding section, we saw that the $V_{cf}$
potential used to describe \ex{11}Be has a significant
influence on the breakup cross section.
A thorough analysis of this influence is therefore
necessary in order to find a convenient way to
constrain the potential choice.
A first step in that direction has been done in \Ref{CBM03b}.
In that previous work, we have studied the influence
of the Pauli-forbidden states on the Coulomb breakup
of \ex{11}Be.
As explained earlier, \ex{11}Be is usually described
by deep potentials that exhibit spurious bound states
besides the adjusted physical levels.
Those unphysical states simulate orbitals occupied by
the neutrons of the core. They are thus forbidden to the
halo neutron by the Pauli principle.
Their presence is usually ignored \cite{KYS94,EBB95,MB99,TS01r,Fal02}.
However, in reaction models, nothing prevents
the transfer of the halo neutron toward one of
those spurious states.
It is therefore interesting to test the influence of those
states upon our calculations.

It is possible to modify a potential in order to remove
one of its bound states \cite{Bay87a}.
The modification consists in a pair of supersymmetric
transformations that keeps all the other spectrum properties
of the potential unchanged.
It means that the supersymmetric partner of the potential
exhibits the same scattering properties (i.e. phase shifts)
and the same bound spectrum (i.e. energy levels)
as the initial potential, except for the bound state
that has been removed.

\begin{figure}
  \includegraphics[height=7cm]{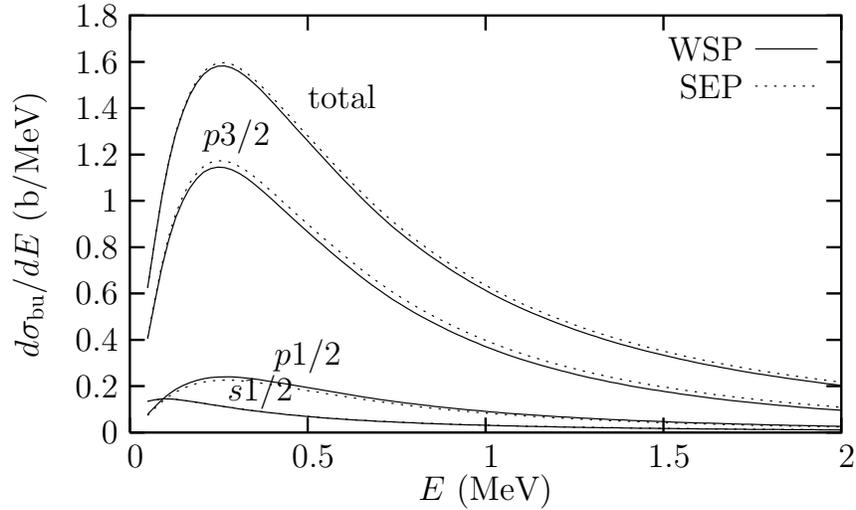}
  \caption{Influence of Pauli-forbidden states upon
the breakup cross section of \ex{11}Be on \ex{208}Pb.
The full lines correspond to the calculation performed
with a Woods-Saxon potential (WSP),
while the dotted lines display the results obtained
with the supersymmetric equivalent potential (SEP).
The latter exhibits the same scattering properties
and the same bound spectrum as the former
but for the unphysical bound states $0s1/2$ and $0p3/2$,
which have been removed.
}\label{f5}
\end{figure}

In \Ref{CBM03b}, we perform two evolution calculations:
one with a usual Woods-Saxon potential (WSP), and one with
its supersymmetric equivalent potential (SEP) in which
both the unphysical $0s1/2$ and $0p3/2$ states have been
suppressed.
The results are shown in \fig{f5}.
The full lines correspond to the breakup cross section
computed with the WSP, and the dotted lines show the
result obtained with the SEP.
Contributions of the major partial waves are pictured as well.
The dependence of the cross section on the potential
is very weak. The difference is only 1\% in the peak region.
Note that the effect differs according to the partial wave.
We observe a slight increase (2.5\%) of the $p3/2$ contribution
when the SEP is used.
The opposite is obtained for the $p1/2$ component: the SEP cross
section is smaller by 5\%.
The $s1/2$ contribution remains practically unchanged.

This result shows that the Pauli-forbidden states in the projectile
spectrum do not play any significant role in the breakup reaction.
They may be ignored.
The use of deep potentials seems therefore fully justified
in such calculations.
Moreover, the tiny difference observed between WSP and SEP
suggests that the discrepancy
between the various \ex{10}Be-n potentials are not directly
due to the differences in energies of their unphysical-states.


\section{Conclusion}
In this talk, we have presented the results of time-dependent
calculations of the nuclear and Coulomb breakup of \ex{11}Be
\cite{CBM03c,CGB04}.
The calculations are performed around 70~MeV/nucleon
in order to compare them to recent experimental
data \cite{Fuk04}.

The description of \ex{11}Be is improved in comparison
to previous works \cite{KYS94,EBB95,MB99,TS01r,Fal02,CBM03c}.
We developed a new \ex{10}Be-n potential that reproduces
not only the bound states of \ex{11}Be but also
its low-lying $\fial^+$ resonance.
This resonance has a significant influence
on the breakup on \ex{12}C \cite{CGB04}:
it induces a narrow peak in the breakup cross section.
The very good agreement obtained with experiment \cite{Fuk04}
confirms the validity of the model, and the ability
of the time-dependent technique to simulate nuclear induced breakup.
The spectrum of \ex{11}Be includes other low-lying resonances,
which should influence the breakup as well.
However, the analysis of their actual effect on the
cross section requires a better description of the
projectile since these resonances cannot be reproduced
with such a simple two-body model.
In the dissociation on \ex{208}Pb, however,
the $\fial^+$ resonance is found to play a rather minor role.
It only induces a small bump in the breakup cross section at the
resonance energy. In this case also, we obtain a rather
good agreement with experimental data \cite{Fuk04}.

In our reaction model, the nuclear interactions between
the projectile and the target are described by optical
potentials \cite{CBM03c}.
In the breakup on \ex{12}C, they are dominant.
In particular, they are found to be responsible for the
strong population of the resonant state, which
causes the peak in the cross section.
The nuclear induced breakup is therefore rather
sensitive to the optical potential choice.
Using different optical potentials leads to
significant variations in the breakup cross section.
However, these variations only affect the
amplitude of the cross section and not its general pattern.
In particular the location and the shape of the resonance peak
remain the same for all choices of optical-potential.
The breakup of \ex{11}Be on \ex{208}Pb is Coulomb dominated.
The nuclear $P$-$T$ interactions are therefore much less significant,
and the cross section is much less dependent
on the optical potentials.

The sensitivity of the model onto
the \ex{11}Be description has also been presented.
We performed time-dependent calculations
of the Coulomb breakup of \ex{11}Be
using various \ex{10}Be-n potentials.
These potentials lead to
cross sections which exhibit the same shape, but
differ by up to 30\% in amplitude.
This variation is not directly related to the
asymptotic normalization constant of the initial
ground state.
Neither is it to the presence of Pauli-forbidden states
in the \ex{11}Be spectrum.
The role played by these unphysical states in the breakup
is indeed negligible \cite{CBM03b}.
Another effect is thus at play here.
Its analysis requires further investigations.
Anyway, because of this significant variation in the
amplitude of the cross section,
using Coulomb breakup as a tool for extracting
spectroscopic factors seems questionable.

From this analysis, it seems that we have now reached
the limit of the simple two-body description
of halo nuclei used in reaction theory.
In order to improve our results
in the nuclear induced breakup of \ex{11}Be
we need a more precise model that reproduces the other
low-lying resonances.
Moreover, the sensitivity of the breakup cross section
to the current \ex{11}Be model
might be too large to extract accurate structure
information.
Therefore, a reaction model including a more precise
description of halo nuclei should be developed in order
to improve the theoretical predictions.


\begin{theacknowledgments}
This text presents research results of the Belgian program P5/07
on interuniversity attraction poles initiated by the Belgian-state
Federal Services for Scientific, Technical and Cultural Affairs.
P.C. acknowledges the support of
the Natural Sciences and Engineering Research Council of Canada (NSERC).
G.G. acknowledges the support of the FRIA (Belgium).
\end{theacknowledgments}


\end{document}